\documentclass{aa}  

\usepackage{graphicx}
\usepackage[varg]{txfonts}

\usepackage{placeins}             
\usepackage{dblfloatfix}         
\usepackage{multirow}
\usepackage[fleqn]{cases}
\usepackage{array}

\usepackage{xcolor}

\usepackage{natbib,twoopt}
\usepackage[breaklinks=true]{hyperref} 
\usepackage{orcidlink}
\bibpunct{(}{)}{;}{a}{}{,}            

\begin{document} 

   \title{Internal constitution of the outer crust of \\ non-accreted neutron stars and magnetars}

   \author{J. Servais\inst{1,2}
          \and
          N. Chamel\inst{1,2} \fnmsep\thanks{Corresponding author:  \email{nicolas.chamel@ulb.be}}
          }

   \institute{Institut d'Astronomie et d'Astrophysique, CP-226, Universit\'e Libre de Bruxelles, 
1050 Brussels, Belgium\\
         \and
             Brussels Laboratory of the Universe, Belgium\\
             }

   \date{Received 29 April 2026 / Accepted 14 May 2026}

   \abstract
   {Determining the internal constitution of the outer crust of magnetars is important for interpreting several of their astrophysical manifestations. In particular, the crustal composition is a key input for simulations of $r$-process nucleosynthesis in giant flare ejecta. 
   However, traditional methods are computationally expensive, limiting their use in large-scale studies. Although faster iterative approaches exist, they are restricted to unmagnetized matter and strongly quantizing magnetic fields, leaving the intermediate field strengths characteristic of observed magnetars without an efficient treatment.}
   {We developed the program \texttt{magcrust} to extend these existing iterative approaches, enabling the rapid computation of the outer-crust composition of cold, non-accreted magnetars over the full range of the magnetic-field strengths inferred for these objects.} 
   {Transitions between adjacent crustal layers are computed by solving approximate equilibrium conditions at the interface. Nuclear abundances and layer depths are estimated from approximate solutions of Einstein’s equations of general relativity.}
   {The performance and accuracy of the program were assessed against detailed numerical calculations. Relative deviations from exact transition properties remain within a few percent, and crustal compositions are well reproduced across 17 nuclear mass tables and 1300 magnetic-field strengths from $10^{13}$ to $10^{16}$ G. 
   Computation times are reduced by factors of $10^3-10^7$ compared to traditional approaches.}
   {This program provides a robust and efficient tool for determining the stratification of magnetars' outer crust over the full range of astrophysically relevant magnetic-field strengths. Its computational speed makes it well suited to systematic calculations, including sensitivity analyses, uncertainty quantification, and ensemble studies.}

   \maketitle

\section{Introduction}

Born in gravitational core-collapse supernova explosions, neutron stars are among the most extreme stellar objects in the Universe. The subclass of magnetars \citep{1992ApJ...392L...9D} is endowed with the strongest magnetic fields known, reaching $10^{14}-10^{15}~\mathrm{G}$ at their surfaces (see, e.g., \citealp{2026enap....3..205R} for a recent review). Magnetars, which include anomalous X-ray pulsars and soft gamma-ray repeaters (SGRs),  are very active isolated neutron stars.
Their persistent X-ray luminosity, for example, far exceeds what can be accounted for by rotational energy loss and implies surface temperatures well above those observed in isolated ordinary neutron stars \citep{2013MNRAS.434..123V}. This excess energy is thought to originate from the evolution of the magnetic field \citep{2016ApJ...833..261B}, which may lead to heat release through either crustal deformations \citep[see, e.g.,][]{2020ApJ...903...40D} or electron captures by the constituting nuclei of the outer crust \citep{2010ApJ...708L..80C, 2021Univ....7..193C}. The interpretation of such observations relies heavily on resolving the structure and equation of state of the outer crust, as its ability to sustain stresses directly depends on its composition \citep{2018MNRAS.480.5511B}, and the heat deposited in the crust by electron captures can only be estimated if the initial distribution of nuclides is known \citep{2022Univ....8..328C}. 
The most spectacular events powered by magnetars are giant flares, very intense outbursts originating from abrupt rearrangements of the magnetic field (see, e.g., \citealp{2025ApJ...980..211B} for a recent review). A substantial fraction of the outer crust can be ejected in these events, liberating a neutron-rich material in conditions favorable for the rapid neutron capture process ($r$-process) responsible for the production of stable and long-lived radioactive nuclei heavier than iron \citep{2007PhR...450...97A, 2011A&A...531A..78G}. Determining the initial composition of the ejected crustal material is essential to predicting the final yields of $r$-process elements, as well as the optical and ultraviolet transient emission powered by their radioactive decay, the so-called novae breves 
\citep{2024MNRAS.528.5323C, 2025ApJ...985..234P}, and the ensuing gamma-ray emission, as observed in the aftermath of the 2004 December giant flare from SGR~1806$-$20 \citep{2025ApJ...984L..29P}. 

Due to the high densities found in the outer crust of neutron stars, ions are fully ionized and immersed in a charge-neutralizing gas of electrons. Nuclei are arranged in a crystal lattice such that the outer crust is structured in solid layers made of progressively more neutron-rich nuclides with increasing depth. The transition to the inner crust is defined by the neutron-drip point, where the pressure is such that neutrons begin to drip out of nuclei \citep{2018ASSL..457..337B}. The composition of the outer crust has traditionally been obtained \citep{1971A&A....14..451T, 1971ApJ...170..299B, 1991ApJ...383..745L} under the cold-catalyzed matter hypothesis \citep{harrison1658, 1965gtgc.book.....H} by identifying the nuclide that minimizes the Gibbs free energy per nucleon for a given pressure and at zero temperature (see, e.g., \citealp{2015A&A...584A.103S, 2016PhRvC..94f5802C, PhysRevC.93.014311, 2017EPJWC.13709001C, 2018MNRAS.481.2994P, PhysRevC.102.065801, 2022PhRvD.105d3017P} for recent calculations of ordinary neutron stars; and \citealp{2015PhRvC..92c5802B, 2019PhRvC..99e5805M, 2023PhRvD.107d3022P, 2024ChPhC..48g4103J, 2025JPhG...52h5201J} for recent calculations of magnetars). The only inputs needed are the masses of all possible nuclides \citep{2013PhRvL.110d1101W}. 

Because not all masses have been measured yet, especially for extremely neutron-rich isotopes (see, e.g., \citealp{2021PrPNP.12003882Y} for a review), theoretical mass tables calculated from nuclear models are required \citep{2003RvMP...75.1021L}. This is especially relevant for magnetars, as their strong magnetic fields, which are not accessible in terrestrial laboratories, can modify the structure and properties of nuclei \citep{2011PhRvC..84d5806P, 2016PhRvC..94c5802S}, in turn altering the composition of the outer crust \citep{2015PhRvC..92c5802B, 2024ChPhC..48g4103J, 2025JPhG...52h5201J}. To avoid missing any thin layers, especially in the deepest region of the outer crust where the most abundant nuclides can be found, 
a sufficiently fine pressure grid is necessary (see, e.g., \citealp{2013IJMSp.349...63K}). This requirement quickly renders the procedure computationally expensive for large-scale statistical studies (such as \citealp{2020PhRvC.101c5804P} in the absence of a magnetic field) and/or systematic calculations across a wide range of magnetic field strengths, as required for the modeling of magnetar crusts. 

A computationally much faster algorithm was presented in \citet{2020PhRvC.101c2801C}, and a program was made publicly available in \citet{2020zndo...3719439C}. The stratification of the outer crust of cold, non-accreting, and unmagnetized neutron stars is determined iteratively, making use of very accurate analytical formulas to calculate matter properties at the transition between adjacent crustal layers. This approach was extended to magnetars in \citet{2020PhRvC.101f5802C}, and another program was released in \citet{2020zndo...3839787C}. However, that program, which still employs very accurate analytical formulas, is limited to extreme magnetic fields, $B \gtrsim 5 \times 10^{16}~\mathrm{G}$, significantly stronger than those observed at the surface of known magnetars. 

In this work, we address these limitations by presenting a new program, \texttt{magcrust}, that is based on the same iterative treatment but allows for more realistic, weaker magnetic fields. Even though analytic solutions are not available in this regime, this method remains computationally much less costly than the traditional minimization. This new program thus enables the efficient and accurate determination of the outer crust composition of cold and isolated neutron stars with arbitrary magnetic-field strengths. It is particularly relevant for magnetars with typical surface fields of $10^{14}-10^{15}~\mathrm{G}$, outside the range treated by the abovementioned programs. 

The general equations governing dense Coulomb crystals are presented in Sect. \ref{sec:general}, and the conditions for transitions between adjacent crustal layers are derived in Sect. \ref{sec:transitions}. Section \ref{sec:structure} addresses the global structure of neutron stars and provides analytical approximations for nuclear abundances and depths marking the transitions between layers. Our numerical method is described in Sect. \ref{sec:numerical}, and the applications and robustness of the program are discussed in Sect. \ref{sec:results}.

\section{Dense Coulomb crystal}
\label{sec:general}

The crustal region considered in this work spans matter densities from the ionization threshold up to the neutron-drip point. Following \citet{2020PhRvC.101c2801C} and \citet{2020PhRvC.101f5802C}, the outer crust is considered to be stratified into pure layers, each composed of a one-component Coulomb crystal embedded in a charge-compensating electron gas. In principle, binary compounds could exist at the interface but their abundance is marginal~\citep{2016PhRvC..94f5802C}. Each layer is therefore described by a nuclear species $(A,Z)$, with mass number $A$ and atomic number $Z$, in thermodynamic equilibrium at a temperature $T$ below the crystallization temperature. For practical purposes, the temperature is set to $T=0$~K \citep{2020A&A...633A.149F}. The electron gas can be accurately treated as a nearly ideal, relativistic, and strongly degenerate Fermi gas. The key equations are summarized here. Further details can be found in \citet{2020PhRvC.101c2801C}, \citet{2020PhRvC.101f5802C}, 
and the references therein. 

In magnetars, the intense magnetic field quantizes transverse electron motion into discrete Landau-Rabi levels. This effect is significant for magnetic fields of strength $B$ above the critical value 
\begin{equation}
B_\mathrm{cr} = \dfrac{m_\mathrm{e}^2c^3}{e \hbar} \approx 4.414 \times 10^{13}~\mathrm{G}
\end{equation}
 at which the electron cyclotron energy becomes comparable to the electron rest-mass energy, with $m_\mathrm{e}$ the electron mass, $c$ the speed of light, $e$ the elementary electric charge, and $\hbar$ the Planck-Dirac constant. Introducing the dimensionless magnetic-field strength
\begin{equation}
  B_* \equiv B/B_\mathrm{cr} ,
\end{equation}
we refer to neutron stars with $B_*<1$ as ordinary neutron stars and those with $B_* \geq 1$ as magnetars. 

The number of Landau-Rabi levels populated by the electrons in a given layer of the crust is dictated by the index of the highest occupied level,
\begin{equation}\label{eq:nu_max}
    \nu_\mathrm{max} = \left\lfloor \dfrac{\gamma_\mathrm{e}^2 - 1}{2B_*} \right\rfloor,
\end{equation}
where $\gamma_\mathrm{e} \equiv \mu_\mathrm{e} / m_\mathrm{e}c^2$ is the electron chemical potential $\mu_\mathrm{e}$ in units of the electron rest-mass energy  and $\left\lfloor \cdot \right\rfloor$ denotes the integer part. As the electron chemical potential varies with the depth in the crust, two adjacent layers can have different $\nu_\mathrm{max}$ for a given magnetic-field strength.
In turn, $\gamma_\mathrm{e}$ is determined by the electron number density, $n_\mathrm{e}$, through a summation over all occupied levels of index $\nu$, 
\begin{equation}
  n_\mathrm{e} = \dfrac{2B_*}{\left(2\pi\right)^2  \lambda_\mathrm{e}^{3}} \sum_{\nu=0}^{\nu_\text{max}} g_\nu x_\mathrm{e}(\nu),
  \label{eq:electron_density}
\end{equation} 
where $\lambda_\mathrm{e} = \hbar/m_\mathrm{e}c$ is the reduced electron Compton wavelength, $g_\nu$ indicates the level degeneracy with $g_\nu = 1$ for $\nu = 0$ and $g_\nu = 2$ for $\nu \geq 1$, and $x_\mathrm{e}(\nu)$ is a dimensionless parameter defined as
\begin{align}
  x_\mathrm{e} (\nu) = \sqrt{\gamma_\mathrm{e}^2 - 1 - 2\nu B_*}~.
\end{align}
The pressure of the electron gas is given by 
\begin{align}
  &P_\mathrm{e} = \dfrac{B_* m_\mathrm{e} c^2}{\left(2\pi\right)^2 \lambda_\mathrm{e}^{3}} \, \sum_{\nu = 0}^{\nu_\mathrm{max}} g_{\nu} \left(1 + 2 \nu B_*\right) \, \psi \left( \dfrac{x_\mathrm{e}(\nu)}{\sqrt{1 + 2\nu B_*}} \right), \label{eq:electron_pressure} \\[10pt]
  &\text{with }\; \psi (x) = x \sqrt{1 + x^2} - \ln \left(x + \sqrt{1 + x^2}\right).
\end{align}

In the most extreme strongly quantizing regime considered in \citet{2020PhRvC.101f5802C}, for which all electrons are confined to the lowest level such that $\nu_\mathrm{max}=0$, Eqs.~(\ref{eq:electron_density}) and (\ref{eq:electron_pressure}) reduce, respectively, to
\begin{align}
  &n_\mathrm{e} = \dfrac{B_*}{2\pi^2 \lambda_\mathrm{e}^{3}} \, x_\mathrm{e} \label{eq:electron_dens_s}, \\[5pt]
  &P_\mathrm{e} = \dfrac{B_* m_\mathrm{e} c^2}{4\pi^2 \lambda_\mathrm{e}^{3}} \, \left[ x_\mathrm{e} \sqrt{1+x_\mathrm{e}^2} - \ln \left(x_\mathrm{e} + \sqrt{1+x_\mathrm{e}^2}\right) \right],\label{eq:pressure_s} 
\end{align}
with $x_\mathrm{e} = \sqrt{\gamma_\mathrm{e}^2 - 1}$. In the opposite limit of a vanishing magnetic field considered in \citet{2020PhRvC.101c2801C}, such that $\nu_\mathrm{max} \rightarrow \infty$, the quantization of electron motion can be ignored. The electron number density and pressure are respectively given by 
\begin{align}
  &n_\mathrm{e} = \dfrac{1}{3\pi^2 \lambda_\mathrm{e}^{3}}\,x_\mathrm{e}^3 ~, \label{eq:electron_dens0} \\[5pt]
  &P_\mathrm{e} = \dfrac{m_\mathrm{e}c^2}{8\pi^2 \lambda_\mathrm{e}^{3}} \left[x_\mathrm{e} \sqrt{1+x_\mathrm{e}^2} \left(\dfrac{2x_\mathrm{e}^2}{3} - 1\right) + \ln \left(x_\mathrm{e} + \sqrt{1 + x_\mathrm{e}^2}\right)\right]. \label{eq:pressure0}
\end{align} 
The main deviation from the ideal electron Fermi gas pressure arises from the lattice pressure, $P_\mathrm{L}$. Ignoring quantum zero-point motion of ions \citep{2009PhRvE..80d6405B}, $P_\mathrm{L}$ takes the same form for magnetars and ordinary neutron stars, and can be generically expressed as 
\begin{align}
    P_\mathrm{L} = \dfrac{C \alpha\hbar c}{3}\, n_\mathrm{e}^{4/3} Z^{2/3}, \label{eq:lattice_pressure}
\end{align}
where $\alpha = e^2/\hbar c$ is the fine structure constant and $C\equiv C_\mathrm{M} \left(4\pi / 3\right)^{1/3}$ is the lattice structure constant with $C_\mathrm{M}<0$ the Madelung constant. For a discussion of higher-order corrections, see, for example, \cite{2011PhRvC..83f5810P} and \cite{2015PhRvD..92b3008C}. 

The matter pressure in any given layer of the outer crust finally consists of two contributions,
\begin{align}
    P(n_\mathrm{e},Z,B_*) = P_\mathrm{e}(n_\mathrm{e},B_*) + P_\mathrm{L}(n_\mathrm{e},Z). \label{eq:total_pressure} 
\end{align}
Note that $P_\mathrm{L}<0$, thus reducing the pressure of the electron Fermi gas.

\section{Transition between adjacent crustal layers}
\label{sec:transitions}

Regardless of whether the neutron star is strongly magnetized or not, the equilibrium composition of the crust at a given pressure ($P$) is found by minimizing the Gibbs free energy per nucleon \citep[see, e.g., Appendix A in][]{2015PhRvD..92b3008C}, 
\begin{align}
  g = \dfrac{M'(A,Z)c^2}{A} + \dfrac{Z}{A} m_\mathrm{e}c^2 \left[\gamma_\mathrm{e} - 1 + \dfrac{4}{3} \, C\alpha \lambda_\mathrm{e} \, n_\mathrm{e}^{1/3} Z^{2/3}\right], \label{eq:gibbs_free_energy}
\end{align}
where $ M'(A,Z)$ is the mass of the nuclide $(A,Z)$, including the rest mass of $Z$ electrons. 
Recalling that $g$ coincides with the baryon chemical potential, $\mu$, the transition from a crustal layer made of nuclei $(A_1, Z_1)$ to a denser layer made of nuclei $(A_2,Z_2)$ is determined by the equilibrium conditions
\begin{align}
    &g(A_1,Z_1,n_{\mathrm{e},1},B_*) = g(A_2,Z_2,n_{\mathrm{e},2},B_*)\equiv \mu_{1\rightarrow2}, \label{eq:gibbs_cond} \\
    &P(n_{\mathrm{e},1},Z_1,B_*) = P(n_{\mathrm{e},2},Z_2,B_*)\equiv P_{1\rightarrow2} . \label{eq:pressure_cond}
\end{align}
While $P_\mathrm{e}$ is a continuous function of $n_\mathrm{e}$, the dependence of $P_\mathrm{L}$ on $Z$ induces a jump in density between the two layers whenever $Z_1\neq Z_2$. However, as $\vert P_\mathrm{L}\vert \ll P_\mathrm{e}$, this discontinuity is typically very small and of order $\alpha$. Setting $n_{\mathrm{e},1}=n_\mathrm{e}$ and $n_{\mathrm{e},2} = n_\mathrm{e} + \delta n_\mathrm{e}$, and expanding both the pressure and the Gibbs free energy per nucleon to first order in $\alpha$, the equilibrium condition~\eqref{eq:gibbs_cond} is approximately given by \citep{2020PhRvC.101c2801C, 2020PhRvC.101f5802C}
\begin{align}
    & \gamma_\mathrm{e} + C\alpha \lambda_\mathrm{e} n_\mathrm{e}^{1/3} F(Z_1, A_1, Z_2, A_2) = \gamma_\mathrm{e}^{1\rightarrow2} \label{eq:layer_transition} , \\
    & F(Z_1, A_1, Z_2, A_2) \equiv \left(\dfrac{4}{3}\dfrac{Z_1^{2/3} Z_1}{A_1} - \dfrac{1}{3}\dfrac{Z_1^{2/3} Z_2}{A_2} - \dfrac{Z_2^{2/3} Z_2}{A_2}\right) \notag \\ 
    &\qquad\qquad\qquad\qquad\qquad \times \left(\dfrac{Z_1}{A_1} - \dfrac{Z_2}{A_2}\right)^{-1} \!\!, \\
    & \gamma_\mathrm{e}^{1\rightarrow2} \equiv \left(\dfrac{M'(A_2,Z_2)c^2} {A_2m_\mathrm{e}c^2} - \dfrac{M'(A_1,Z_1)c^2}{A_1m_\mathrm{e}c^2}\right) \left(\dfrac{Z_1}{A_1} - \dfrac{Z_2}{A_2}\right)^{-1} + 1  ~.
\end{align}
For the singular case $Z_1/A_1=Z_2/A_2$, the equilibrium condition can be directly solved for $n_\mathrm{e}$:  
\begin{align}
    &n_\mathrm{e} =\left[ \dfrac{1}{C \alpha \lambda_\mathrm{e} (Z_1^{2/3} - Z_2^{2/3})} \dfrac{A_1}{Z_1} \left( \dfrac{M'(A_2,Z_2)c^2} {A_2m_\mathrm{e}c^2} - \dfrac{M'(A_1,Z_1)c^2}{A_1m_\mathrm{e}c^2} \right) \right]^3. \label{eq:layer_transition_y1y2}
\end{align}

The transition pressure, $P_{1\rightarrow2}$, and associated baryon chemical potential, $\mu_{1\rightarrow2}$, are computed from Eqs.~(\ref{eq:total_pressure}) and (\ref{eq:gibbs_free_energy}), respectively, with $Z=Z_1$, $A=A_1$, and using the solution $\gamma_\mathrm{e}$ of Eq. (\ref{eq:layer_transition}) or calculated from Eq.~\eqref{eq:electron_density} using Eq. (\ref{eq:layer_transition_y1y2}). Finally, the maximum mean nucleon number density, $\overline{n}_1^\mathrm{max}$, up to which the nuclide $(A_1,Z_1)$ is present and the minimum mean nucleon number density, $\overline{n}_2^\mathrm{min}$, at which the nuclide $(A_2,Z_2)$ appears are respectively given by
\begin{align}
  \overline{n}_1^\mathrm{max} &= \dfrac{A_1}{Z_1} \, n_\mathrm{e} , \\
  \overline{n}_2^\mathrm{min} &= \dfrac{A_2}{Z_2} \, n_\mathrm{e} \, \Biggl\{ 1 + \dfrac{C\alpha \hbar c}{3} n_\mathrm{e}^{1/3} \left(Z_1^{2/3} - Z_2^{2/3}\right) \left[ \dfrac{\partial P_\mathrm{e}(n_\mathrm{e},B_*)}{\partial n_\mathrm{e}} \right]^{-1}\Biggr\}. \label{eq:min_dens}
\end{align}

The neutron-drip transition delimiting the bottom of the outer crust occurs when the Gibbs free energy per nucleon $g(A_2,Z_2,n_{\mathrm{e},2},B_*)$ exceeds the neutron rest-mass energy $m_\mathrm{n}c^2$, where $m_\mathrm{n}$ is the neutron mass \citep{2015PhRvC..91e5803C, 2015PhRvC..91f5801C}. The equilibrium condition becomes
\begin{align}
  & \gamma_\mathrm{e} + \dfrac{4}{3} C \alpha \lambda_\mathrm{e} n_\mathrm{e}^{1/3} Z_1^{2/3} = \gamma_\mathrm{e}^\mathrm{drip} , \label{eq:drip_transition} \\[5pt]
  & \gamma_\mathrm{e}^\mathrm{drip} \equiv \dfrac{-M'(A_1,Z_1)c^2 + A_1 m_\mathrm{n}c^2}{Z_1 m_\mathrm{e}c^2} + 1 .
\end{align}
This equation can be solved in the same manner as Eq.~(\ref{eq:layer_transition}), replacing $\gamma_\mathrm{e}^{1\rightarrow2}$ by $\gamma_\mathrm{e}^\mathrm{drip}$ and $F(Z_1,A_1,Z_2,A_2)$ by $(4/3)Z_1^{2/3}$.

The equilibrium conditions marking the interface between crustal layers formally take the same form for both ordinary neutron stars and magnetars. However, they implicitly depend on the magnetic field through the relation~\eqref{eq:electron_density} between $\gamma_\mathrm{e}$ and $n_\mathrm{e}$, and the nuclear masses for sufficiently strong $B_*$.

\section{Global structure and nuclear abundances}
\label{sec:structure}

The strong magnetic fields present in magnetars affect not only their internal composition but also their global structure. Accordingly, the computation of the crustal structure and nuclear abundances formally requires the simultaneous resolution of Einstein and Maxwell equations \citep{2019PhRvC..99e5811C}. However, such calculations rely on assumptions about the unknown geometry of the magnetic fields. Given these uncertainties, and since the influence of the field on the stellar radius has been shown to remain below 1–2\% for fields $B \lesssim 4 \times 10^{17}~\mathrm{G}$ \citep{2019A&A...627A..61G}, we adopted the analytical expressions derived for unmagnetized neutron stars by \citet{2020PhRvC.101c2801C} and applied them to both ordinary neutron stars and magnetars. The relative nuclear abundance, $\xi_i$, of a crustal layer made of nuclei $(A_i,Z_i)$ is approximated as
\begin{align}
    \xi_i = \frac{P_{i\rightarrow i+1}-P_{i-1\rightarrow i}}{P_\mathrm{drip}} ,
\end{align}
where $P_\mathrm{drip}$ is the matter pressure at the neutron-drip point \citep{2020PhRvC.101c2801C}. The relative proper depth, $z_i/z_\mathrm{drip}$,  at the bottom of this crustal layer is
\begin{align}
    \frac{z_i}{z_\mathrm{drip}} \approx \dfrac{\left( \mu_{i\rightarrow i+1} / \mu_s \right)^2 - 1}{\left( m_\mathrm{n}c^2/\mu_\mathrm{s} \right)^2 - 1},
\end{align}
where $z_\mathrm{drip}$ is the proper depth at the neutron-drip point and $\mu_\mathrm{s}$ the baryon chemical potential at the stellar surface. The latter is computed differently for ordinary neutron stars and magnetars, as discussed in \citet{2020PhRvC.101c2801C} and \citet{2020PhRvC.101f5802C}, respectively.

\section{Numerical implementation}
\label{sec:numerical}

Instead of minimizing the Gibbs free energy per nucleon on a pressure grid as in the traditional method, the iterative approach we adopted focuses on the transitions between layers. Given a crustal layer composed of nuclei $(A_1,Z_1)$, the composition of the layer beneath is obtained by solving Eq. (\ref{eq:layer_transition}) or Eq. (\ref{eq:layer_transition_y1y2}) for each candidate nuclide $(A_2,Z_2)$ in a given mass table. Among these, the nuclide that yields the lowest transition pressure ($P_{1\rightarrow2}$) while satisfying the mechanical stability condition $\overline{n}_1^\mathrm{max} \leq \overline{n}_2^\mathrm{min}$ is selected. Assuming a surface layer composed of \element[][56]{Fe}, this procedure is repeated iteratively to determine the sequence of equilibrium nuclides throughout the crust up to the neutron-drip point. Once the composition is found, the relative depths of the layers and the relative nuclear abundances can be readily calculated. 

This method, therefore, revolves around solving the equilibrium condition for an arbitrary number of populated Landau-Rabi levels.

\subsection{Equilibrium condition for arbitrary Landau-Rabi filling}

Equation~\eqref{eq:layer_transition} admits an explicit solution in two limiting cases. 
In the absence of a magnetic field ($\nu_\mathrm{max} \rightarrow \infty$), the electron number density is given by Eq.~(\ref{eq:electron_dens0}). 
As described in \citet{2020PhRvC.101c2801C}, Eq.~(\ref{eq:layer_transition}) then reduces to a quadratic polynomial equation that can be solved analytically. In the strongly quantizing regime ($\nu_\mathrm{max}=0$), the electron number density is given by Eq.~(\ref{eq:electron_dens_s}). Under the ultra-relativistic approximation $\gamma_\mathrm{e} \gg 1$, the equilibrium condition (\ref{eq:layer_transition}) reduces to a cubic polynomial equation and is therefore also amenable to an analytic solution, as shown in \citet{2020PhRvC.101f5802C}. 
Equation~(\ref{eq:layer_transition_y1y2}) is solved analytically for $\gamma_\mathrm{e}$, using Eq.~(\ref{eq:electron_dens0}) or Eq.~(\ref{eq:electron_dens_s}) in these corresponding limits.

Between these two limiting cases, an arbitrary number of Landau-Rabi levels is populated, complexifying the expressions for the electron number density and pressure.
When two Landau-Rabi levels are occupied, i.e., $\nu_\mathrm{max}=1$, they become
\begin{align}
        & n_\mathrm{e} = \dfrac{B_*}{2\pi^2 \lambda_\mathrm{e}^{3}} \left[\sqrt{\gamma_\mathrm{e}^2 - 1} + 2 \sqrt{\gamma_\mathrm{e}^2 - 1 - 2B_*}\right] , \label{eq:ne_int}
\end{align}
\begin{align}
    &P = \dfrac{B_* m_\mathrm{e} c^2}{4\pi^2 \lambda_\mathrm{e}^{3}} \left[ \psi \left(\sqrt{\gamma_\mathrm{e}^2 -1}\right) + 2\,(1 + 2B_*) \, \psi \left(\dfrac{\sqrt{\gamma_\mathrm{e}^2 - 1 - 2B_*}}{\sqrt{1 + 2B_*}}\right) \right. \notag \\
                &\qquad\quad + \left. \dfrac{C\alpha}{3} \left(\dfrac{4B_*}{\pi^2}\right)^{\!1/3} Z_1^{2/3} \, \left(\sqrt{\gamma_\mathrm{e}^2 - 1} + 2 \sqrt{\gamma_\mathrm{e}^2 - 1 - 2B_*}\right)^{\!4/3} \right]. \notag \\
\end{align}
The equilibrium condition, Eq. (\ref{eq:layer_transition}), is then given by
\begin{align}
        \gamma_\mathrm{e} & + C\alpha \left\lbrace \dfrac{B_*}{2\pi^2}  \left[ \sqrt{\gamma_\mathrm{e}^2 - 1} + 2 \sqrt{\gamma_\mathrm{e}^2 - 1 - 2B_*} \right] \right\rbrace ^{\!1/3} \notag \\ 
    & \qquad\qquad\qquad\qquad\qquad \times F(Z_1, A_1, Z_2, A_2) = \gamma_\mathrm{e}^{1\rightarrow2},
        \label{eq:eq_cond_int}
\end{align}
and Eq. (\ref{eq:layer_transition_y1y2}) by
\begin{align}
    &\sqrt{\gamma_\mathrm{e}^2 - 1} + 2 \sqrt{\gamma_\mathrm{e}^2 - 1 - 2B_*} = \notag \\
    & \quad \dfrac{2\pi^2}{B_*} \left\lbrace \dfrac{1}{C \alpha (Z_1^{2/3} - Z_2^{2/3})} \dfrac{A_1}{Z_1} \left( \dfrac{M'(A_2,Z_2)c^2} {A_2m_\mathrm{e}c^2} - \dfrac{M'(A_1,Z_1)c^2}{A_1m_\mathrm{e}c^2} \right) \right\rbrace^3.
\end{align}
The coefficient $2B_*/\gamma_\mathrm{e}^2$ being of order unity, no approximation can be made to derive analytic solutions. These equations must therefore be solved numerically.

Writing the explicit expressions for higher $\nu_\mathrm{max}$ is tedious, and evaluating the sums in Eqs.~\eqref{eq:electron_density} and \eqref{eq:electron_pressure} for each possible transition quickly becomes computationally very time-consuming as $\nu_\mathrm{max}$ is increased. For weakly quantizing magnetic fields such that $\nu_\mathrm{max}\gg 1$, it is more convenient to write the electron number density and pressure using the expansions derived in \citet{2001NuPhB.612..492D}, respectively,
\begingroup
\allowdisplaybreaks
\begin{align}
  n_\mathrm{e} \approx \dfrac{1}{2\pi^2 \lambda_\mathrm{e}^{3}} & \left[\dfrac{2}{3}\left(\gamma_\mathrm{e}^2-1\right)^{3/2} \right. \notag \\
  & \quad \left. + (2B_*)^{3/2} \zeta\left(\dfrac{-1}{2}, \left\lbrace \dfrac{\gamma_\mathrm{e}^2-1}{2B_*} \right\rbrace\right)  + \dfrac{B_*^2}{6\sqrt{\gamma_\mathrm{e}^2 - 1}}\right], \label{eq:electron_dens_w}
\end{align}
\begin{align}
  P \approx& \dfrac{m_\mathrm{e}c^2}{4\pi^2 \lambda_\mathrm{e}^{3}}  \left\lbrace \dfrac{1}{2} \left(1 - 2B_* + \dfrac{2B_*^2}{3}\right) \ln \left(\dfrac{\gamma_\mathrm{e} + \sqrt{2B_* + \gamma_\mathrm{e}^2 - 1}}{1 + \sqrt{2B_*}}\right) \right. \notag \\
        &\qquad\qquad - \dfrac{1}{2} \left(\gamma_\mathrm{e}\sqrt{2B_* + \gamma_\mathrm{e}^2 - 1} - \sqrt{2B_*}\right) \notag \\
    &\qquad\qquad + \dfrac{1}{3} \left(\gamma_\mathrm{e}\left( 2B_* + \gamma_\mathrm{e}^2 - 1\right)^{3/2} - \left( 2B_* \right)^{3/2}\right) \notag\\
        &\qquad\qquad + B_* \left(\mathrm{arccosh} \left(\gamma_\mathrm{e}\right) - \gamma_\mathrm{e} \sqrt{\gamma_\mathrm{e}^2 -1}\right) \notag \\
    &\qquad\qquad - \left(2B_*\right)^{5/2} \int_0^{+\infty} \dfrac{\tilde{\zeta}_3(-1/2,q+1)}{\sqrt{1 + 2qB_*}} \text{d}q \notag \\
        &\qquad\qquad + \dfrac{2}{3} \dfrac{\left(2B_*\right)^{5/2}}{\gamma_\mathrm{e}}\, \zeta\left(\dfrac{-3}{2}, \left\lbrace \dfrac{\gamma_\mathrm{e}^2 -1}{2B_*}\right\rbrace \right) \notag \\
    &\qquad\qquad + \dfrac{2}{15} \dfrac{\left( 2B_* \right)^{7/2}}{\gamma_\mathrm{e}^3}\, \zeta\left(\dfrac{-5}{2}, \left\lbrace \dfrac{\gamma_\mathrm{e}^2 -1}{2B_*}\right\rbrace \right) + \dfrac{1}{240} \left(\dfrac{B_*}{\gamma_\mathrm{e}}\right)^{\!4} \notag \\
        &\qquad\qquad + 4B_*^2 \int_0^1 \zeta\left(\dfrac{-1}{2},q\right) \zeta\left(\dfrac{1}{2}, q + \dfrac{1}{2B_*}\right) \text{d}q \notag \\
        &\qquad\qquad + \; \dfrac{2}{3} \left(\dfrac{1}{2\pi^2}\right)^{1/3} C\alpha Z_1^{2/3} \left[\dfrac{2}{3}\left(\gamma_\mathrm{e}^2-1\right)^{3/2} \right. \notag \\
    &\qquad\qquad + \left.\left. (2B_*)^{3/2} \zeta\left(\dfrac{-1}{2}, \left\lbrace \dfrac{\gamma_\mathrm{e}^2-1}{2B_*} \right\rbrace\right) + \dfrac{B_*^2}{6\sqrt{\gamma_\mathrm{e}^2 - 1}}\right]^{4/3} \right\rbrace , \label{eq:pressure_w} 
\end{align}
\endgroup
where $\left\lbrace\cdot \right\rbrace$ denotes the fractional part, and $\zeta(z,q)$ is the Hurwitz zeta function,
\begin{equation}
  \zeta(z,q) = \sum_{\upsilon = 0}^{+\infty} \dfrac{1}{\left(\upsilon + q\right)^z},
\end{equation}
valid for $\text{Re}[z] > 1$ and $q \neq 0,-1,-2,...~$. It can be extended to other $z\neq1$ by analytic continuation. The function $\tilde{\zeta}_3(z,q)$ is defined as
\begin{equation}
  \tilde{\zeta}_3(z,q) = \zeta(z,q) - \dfrac{1}{z-1}\, q^{-z+1} - \dfrac{1}{2}\, q^{-z} - \dfrac{z}{12} \, q^{-z-1} .
\end{equation}
In the limit of a vanishing magnetic field, Eq.~\eqref{eq:electron_dens_w} reduces to Eq.~\eqref{eq:electron_dens0}, while Eq.~\eqref{eq:pressure_w} reduces to the sum of Eqs.~\eqref{eq:pressure0} and \eqref{eq:lattice_pressure}. 
They remain remarkably accurate up to the onset of the strongly quantizing regime, where they show relative deviations below $0.1\%$ and $0.02\%,$ respectively, compared to the exact results (see \citealp{2022Univ....8..328C} for detailed numerical comparisons). We used Eqs.~\eqref{eq:electron_dens_w} and \eqref{eq:pressure_w} in our program for every transition with $\nu_\mathrm{max}\geq 2$. 

As the electrostatic correction in Eq. (\ref{eq:layer_transition}) is small for weakly quantizing magnetic fields, \citet{2022Univ....8..328C} suggested to approximate $\gamma_\mathrm{e}$ in this regime by its solution in the absence of a magnetic field, and substitute it in Eqs.~\eqref{eq:electron_dens_w} and \eqref{eq:pressure_w} to obtain the electron density ($n_\mathrm{e}$) and pressure ($P$). However, we found that this method could cause numerical instabilities at the boundary between the weakly and strongly quantizing regimes. The equilibrium conditions are therefore solved numerically instead. Equation~(\ref{eq:layer_transition_y1y2}) has only one solution and can be solved easily. Equation~(\ref{eq:layer_transition}), however, can have multiple and is solved following the procedure outlined below.

\subsection{Numerical solutions of the equilibrium condition}

\begin{figure}
    \includegraphics[width=\hsize-0.3cm]{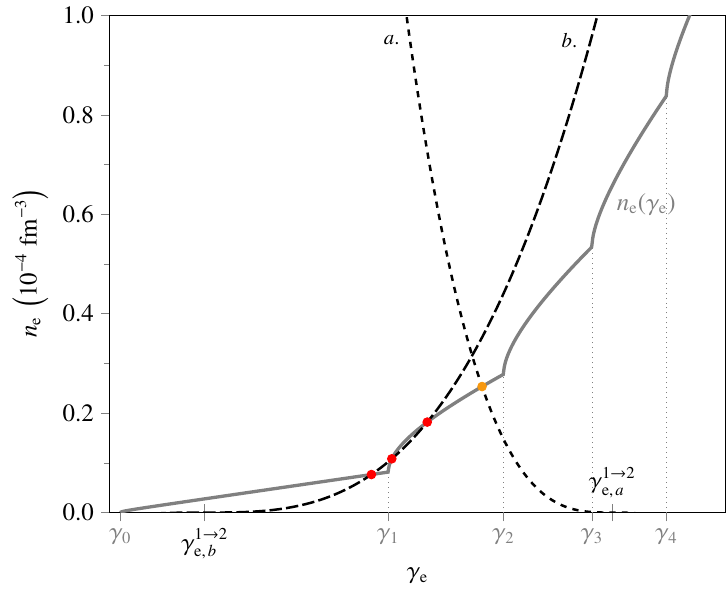}
    \caption{Electron number density ($n_\mathrm{e}$; solid line) and two illustrative forms of $n_\mathrm{e}^{1\rightarrow2}$ (dashed lines labeled $a$ and $b$) as functions of the electron chemical potential ($\gamma_\mathrm{e}$). The threshold energies ($\gamma_k$) are indicated along the horizontal axis. The colored dots mark solutions of Eq.~(\ref{eq:eq_rewritten}).}
     \label{fig:diagram}
\end{figure}

Let us introduce
\begin{equation}
    n_\mathrm{e}^{1\rightarrow2} (\gamma_\mathrm{e}) \equiv \left[\dfrac{\gamma_\mathrm{e}^{1\rightarrow2} - \gamma_\mathrm{e}}{C\alpha \lambda_\mathrm{e} F(Z_1,A_1,Z_2,A_2)}\right]^3 ,
\end{equation}
such that the equilibrium condition~\eqref{eq:layer_transition} can be written as 
\begin{equation}
    n_\mathrm{e}(\gamma_\mathrm{e})=n_\mathrm{e}^{1\rightarrow2}(\gamma_\mathrm{e}). \label{eq:eq_rewritten}
\end{equation}
For Eq. (\ref{eq:eq_rewritten}) to be satisfied, $n_\mathrm{e}^{1\rightarrow2}(\gamma_\mathrm{e})$ must be obviously positive. Since $C<0$, this is only true if
\begin{align}
    & \gamma_\mathrm{e}^{1\rightarrow2} \geq \gamma_\mathrm{e} \text{ and } F(Z_1,A_1,Z_2,A_2) < 0 \text{, or } \\
    & \gamma_\mathrm{e}^{1\rightarrow2} \leq \gamma_\mathrm{e} \text{ and } F(Z_1,A_1,Z_2,A_2) > 0. \label{eq:mathcal_cond}
\end{align}
In the first case, $n_\mathrm{e}^{1\rightarrow2}(\gamma_\mathrm{e})$ is monotonically decreasing while $n_\mathrm{e}(\gamma_\mathrm{e})$ is monotonically increasing (see curve $a$ in Fig.~\ref{fig:diagram}). Equation (\ref{eq:eq_rewritten}) therefore has at most one solution (marked as the orange dot). For this solution to be physical, we further require $\gamma_\mathrm{e} \geq 1$. In the second case, both $n_\mathrm{e}(\gamma_\mathrm{e})$ and $n_\mathrm{e}^{1\rightarrow2}(\gamma_\mathrm{e})$ are monotonically increasing (see curve $b$ in Fig.~\ref{fig:diagram}). Since $\gamma_\mathrm{e}^{1\rightarrow2}$ is the inflection point as well as the only root of $n_\mathrm{e}^{1\rightarrow2}$, $n_\mathrm{e}(\gamma_\mathrm{e})$ and $n_\mathrm{e}^{1\rightarrow2}(\gamma_\mathrm{e})$ have opposite concavities. As follows from Eq.~\eqref{eq:electron_density}, the derivative of $n_\mathrm{e}$ with respect to $\gamma_\mathrm{e}$ diverges for values of $\gamma_\mathrm{e}$ given by
\begin{equation}
    \gamma_k \equiv \sqrt{1 + 2kB_*} \text{ with } k=0,1,2,...
,\end{equation}
which corresponds to the threshold energies for the complete filling of Landau-Rabi levels. At these points, $n_\mathrm{e}(\gamma_\mathrm{e})$ exhibits kinks (see the solid curve in Fig.~\ref{fig:diagram}). Within each subinterval $\left]\gamma_k, \gamma_{k+1}\right]$, however, $n_\mathrm{e}(\gamma_\mathrm{e})$ is strictly convex while $n_\mathrm{e}^{1\rightarrow2}(\gamma_\mathrm{e})$ remains strictly concave. Consequently, Eq. (\ref{eq:eq_rewritten}) admits at most two solutions on each such interval \citep[see, e.g.,][]{Boyd_Vandenberghe_2004}. This behavior is illustrated in Fig.~\ref{fig:diagram}, with Eq.~(\ref{eq:eq_rewritten}) admitting one solution in the subinterval $]\gamma_0,\gamma_1]$, and two in $]\gamma_1,\gamma_2]$ (marked as red dots).
 
For each nuclide candidate $(A_2,Z_2)$, the program identifies all subintervals with possible intersections of the curves $n_\mathrm{e}(\gamma_\mathrm{e})$ and $n_\mathrm{e}^{1\rightarrow2}(\gamma_\mathrm{e})$, and numerically solves Eq.~(\ref{eq:eq_rewritten}) with a standard root-finding algorithm in these intervals. From the solutions $\gamma_\mathrm{e}$ thus found, the corresponding transition pressure ($P_{1\rightarrow2}$) and the mean nucleon number densities ($\overline{n}_1^\mathrm{max}$ and $\overline{n}_2^\mathrm{min}$) are computed. If the magnetic field is sufficiently weak or, on the contrary, strongly quantizing, the analytic solutions corresponding to $\nu_\mathrm{max}\rightarrow +\infty$ and $\nu_\mathrm{max}=0$ are used instead. 

\begin{table*}[!ht]
\caption{Stratification of the outer crust of a magnetar with $B_*=100$.}
\label{tab:strat}   
\renewcommand{\arraystretch}{1.15}
\centering                          
\begin{tabular}{c c c c c c c c c c c c c}     
\hline\hline                 
\rule[0em]{0pt}{1.1em}$Z_1$ & $A_1$ & $Z_2$ & $A_2$ & $\nu_\mathrm{max}$ & $x_\mathrm{e}$ & $\overline{n}_1^\mathrm{max}$ & $\overline{n}_2^\mathrm{min}$ & $P_{1\rightarrow2}$ & $\gamma_\mathrm{e}^{1\rightarrow2}$ & $\mu_{1\rightarrow2}$ & $\xi_1$ & $z_1/z_\mathrm{drip}$ \\ 
\rule[-0.4em]{0pt}{0em}& & & & & & {\footnotesize{$(\mathrm{fm}^{-3})$}} & {\footnotesize{$(\mathrm{fm}^{-3})$}} & {\footnotesize{$(\mathrm{MeV~fm}^{-3})$}} & & {\footnotesize{$(\mathrm{MeV})$}} & & \\ 
\hline 
\rule[0em]{0pt}{1.1em}26 & 56 & 28 & 62 & 0 & 1.50 & $2.84 \times 10^{-7}$ & $2.92 \times 10^{-7}$ & $2.96 \times 10^{-8}$ & 1.89 & 930.5 & $6.10 \times 10^{-5}$ & 0.017 \\
28 & 62 & 28 & 64 & 0 & 5.19 & $1.01 \times 10^{-6}$ & $1.04 \times 10^{-6}$ & $5.41 \times 10^{-7}$ & 4.90 & 931.3 & $1.05 \times 10^{-3}$ & 0.101 \\
28 & 64 & 36 & 86 & 0 & 8.35 & $1.68 \times 10^{-6}$ & $1.76 \times 10^{-6}$ & $1.47 \times 10^{-6}$ & 9.33 & 932.0 & $1.92 \times 10^{-3}$ & 0.175 \\
36 & 86 & 34 & 84 & 0 & 11.1 & $2.33 \times 10^{-6}$ & $2.40 \times 10^{-6}$ & $2.62 \times 10^{-6}$ & 10.0 & 932.6 & $2.37 \times 10^{-3}$ & 0.236 \\
34 & 84 & 32 & 82 & 1 & 16.8 & $7.58 \times 10^{-6}$ & $7.86 \times 10^{-6}$ & $7.35 \times 10^{-6}$ & 15.3 & 933.7 & $9.74 \times 10^{-3}$ & 0.359 \\
32 & 82 & 30 & 80 & 2 & 22.3 & $1.73 \times 10^{-5}$ & $1.80 \times 10^{-5}$ & $1.97 \times 10^{-5}$ & 20.5 & 934.8 & $2.54 \times 10^{-2}$ & 0.474 \\
30 & 80 & 28 & 78 & 3 & 28.2 & $3.38 \times 10^{-5}$ & $3.53 \times 10^{-5}$ & $4.71 \times 10^{-5}$ & 26.0 & 935.8 & $5.63 \times 10^{-2}$ & 0.591 \\
28 & 78 & 44 & 126 & 6 & 35.3 & $7.24 \times 10^{-5}$ & $7.53 \times 10^{-5}$ & $1.13 \times 10^{-4}$ & 47.8 & 937.1 & $1.36 \times 10^{-1}$ & 0.727 \\
44 & 126 & 42 & 124 & 6 & 36.7 & $8.27 \times 10^{-5}$ & $8.52 \times 10^{-5}$ & $1.30 \times 10^{-4}$ & 33.1 & 937.3 & $3.49 \times 10^{-2}$ & 0.751 \\
42 & 124 & 40 & 122 & 8 & 42.0 & $1.28 \times 10^{-4}$ & $1.32 \times 10^{-4}$ & $2.22 \times 10^{-4}$ & 38.0 & 938.2 & $1.90 \times 10^{-1}$ & 0.846 \\
40 & 122 & 38 & 120 & 9 & 44.6 & $1.58 \times 10^{-4}$ & $1.63 \times 10^{-4}$ & $2.83 \times 10^{-4}$ & 40.6 & 938.6 & $1.26 \times 10^{-1}$ & 0.892 \\
38 & 120 & 38 & 122 & 12 & 49.8 & $2.30 \times 10^{-4}$ & $2.34 \times 10^{-4}$ & $4.40 \times 10^{-4}$ & 47.3 & 939.4 & $3.24 \times 10^{-1}$ & 0.980 \\
38 & 122 & 38 & 124 & 12 & 50.9 & $2.47 \times 10^{-4}$ & $2.51 \times 10^{-4}$ & $4.80 \times 10^{-4}$ & 48.3 & 939.5 & $8.18 \times 10^{-2}$ & 0.998 \\
\rule[-0.4em]{0pt}{0em}38 & 124 & -  & - & 13 & 51.1 & $2.55 \times 10^{-4}$ & - & $4.85 \times 10^{-4}$ & 48.5 & 939.6 & $1.11 \times 10^{-2}$ & 1.00 \\
\hline 
\end{tabular}
\tablefoot{
This composition was computed using experimental data from AME2016 supplemented by the HFB-27 nuclear mass model. Each line lists the properties of the transition from a layer composed of the nuclide $(A_1,Z_1)$ to a layer composed of the nuclide $(A_2,Z_2)$. See text for details.
}
\end{table*}

\section{Applications and tests of the program}
\label{sec:results}

\begin{table*}[!hb]
\caption{Relative deviations $100 (q - q_\mathrm{exact})/q_\mathrm{exact}$ (\%) between the approximate crustal properties ($q$) listed in Table \ref{tab:strat} and their exact values ($q_\mathrm{exact}$).}
\label{tab:errors}   
\renewcommand{\arraystretch}{1.15}
\centering                          
\begin{tabular}{c c c c c c c c c c}     
\hline\hline                 
\rule[-0.5em]{0pt}{1.6em}$Z_1$ & $A_1$ & $\nu_\mathrm{max}$ & $x_\mathrm{e}$ & $\overline{n}_1^\mathrm{max}$ & $\overline{n}_2^\mathrm{min}$ & $P_{1\rightarrow2}$ & $\mu_{1\rightarrow2}$ & $\xi_1$ & $z_1/z_\mathrm{drip}$ \\  
\hline                      
\rule[0em]{0pt}{1.1em}26 & 56 & 0 & $-4.2\times 10^{-1}$ & $-4.2\times 10^{-1}$ & $-4.3\times 10^{-1}$ & -1.1 & $-1.2\times 10^{-4}$ & -1.1 & $-7.4\times 10^{-1}$ \\
28 & 62 & 0 & $4.7\times 10^{-2}$ & $4.7\times 10^{-2}$ & $4.7\times 10^{-2}$ & $1.0\times 10^{-1}$ & $5.8\times 10^{-5}$ & $1.7\times 10^{-1}$ & $5.8\times 10^{-2}$ \\
28 & 64 & 0 & $-1.9\times 10^{-2}$ & $-1.9\times 10^{-2}$ & $-2.4\times 10^{-2}$ & $-3.9\times 10^{-2}$ & $-3.6\times 10^{-5}$ & $-1.2\times 10^{-1}$ & $-2.1\times 10^{-2}$ \\
36 & 86 & 0 & $1.4\times 10^{-2}$ & $1.4\times 10^{-2}$ & $1.5\times 10^{-2}$ & $2.9\times 10^{-2}$ & $3.5\times 10^{-5}$ & $1.2\times 10^{-1}$ & $1.5\times 10^{-2}$ \\
34 & 84 & 1 & $8.1\times 10^{-4}$ & $1.8\times 10^{-3}$ & $6.1\times 10^{-3}$ & $2.8\times 10^{-3}$ & $2.9\times 10^{-6}$ & $-1.2\times 10^{-2}$ & $8.2\times 10^{-4}$ \\
32 & 82 & 2 & $8.2\times 10^{-4}$ & $1.9\times 10^{-3}$ & $6.2\times 10^{-3}$ & $3.1\times 10^{-3}$ & $3.7\times 10^{-6}$ & $3.3\times 10^{-3}$ & $8.1\times 10^{-4}$ \\
30 & 80 & 3 & $5.9\times 10^{-4}$ & $1.2\times 10^{-3}$ & $4.0\times 10^{-3}$ & $2.2\times 10^{-3}$ & $3.3\times 10^{-6}$ & $1.6\times 10^{-3}$ & $5.7\times 10^{-4}$ \\
28 & 78 & 6 & $5.9\times 10^{-2}$ & $1.9\times 10^{-1}$ & $1.4\times 10^{-1}$ & $2.3\times 10^{-1}$ & $3.9\times 10^{-4}$ & $4.0\times 10^{-1}$ & $5.5\times 10^{-2}$ \\
44 & 126 & 6 & $8.2\times 10^{-4}$ & $2.0\times 10^{-3}$ & $6.9\times 10^{-3}$ & $3.2\times 10^{-3}$ & $5.5\times 10^{-6}$ & -1.5 & $7.4\times 10^{-4}$ \\
42 & 124 & 8 & $7.9\times 10^{-4}$ & $1.9\times 10^{-3}$ & $6.8\times 10^{-3}$ & $3.2\times 10^{-3}$ & $5.9\times 10^{-6}$ & $3.0\times 10^{-3}$ & $7.1\times 10^{-4}$ \\
40 & 122 & 9 & $7.4\times 10^{-4}$ & $1.7\times 10^{-3}$ & $6.3\times 10^{-3}$ & $2.9\times 10^{-3}$ & $5.6\times 10^{-6}$ & $2.2\times 10^{-3}$ & $6.5\times 10^{-4}$ \\
38 & 120 & 12 & $5.9\times 10^{-8}$ & $1.7\times 10^{-7}$ & $1.7\times 10^{-7}$ & $2.4\times 10^{-7}$ & $4.8\times 10^{-10}$ & $-5.3\times 10^{-3}$ & $5.0\times 10^{-8}$ \\
38 & 122 & 12 & $5.0\times 10^{-8}$ & $1.2\times 10^{-7}$ & $1.2\times 10^{-7}$ & $2.0\times 10^{-7}$ & $4.1\times 10^{-10}$ & $-2.2\times 10^{-7}$ & $4.2\times 10^{-8}$ \\
\rule[-0.4em]{0pt}{0em}38 & 124 & 13 & $4.3\times 10^{-8}$ & $2.8\times 10^{-6}$ & - & $-5.7\times 10^{-7}$ & 0 & $-1.8\times 10^{-5}$ & 0 \\
\hline 
\end{tabular}
\end{table*}

\subsection{Code setup and physical inputs}
The only input required to run the program is a nuclear mass table listing the atomic number ($Z$), mass number ($A$), and nuclear mass (in $\mathrm{MeV}$) of all possible nuclei. The masses $M'(A,Z)$ are then computed by adding the rest mass of $Z$ electrons. Mass tables that include the influence of the magnetic field on nuclear structure can be used without any modification of the program. The strength of the magnetic field of the star 
must be specified at runtime on the command line, in units of $B_*$. By default, the analytic solutions for unmagnetized neutron stars are used for $B_*<1$. This threshold can be adjusted through an environment variable, although increasing it may incur significant errors in the crust composition. 

Values for the fundamental constants used in the program are taken from the 2018 CODATA database \citep{2021RvMP...93b5010T}. Nuclei are considered to form a body-centered cubic lattice independently of the magnetic field strength \citep{2012PhRvC..86e5804C, 2016Ap&SS.361..256K}. By default, the Madelung constant is taken from \citet{2001PhRvE..64e7402B}. It can be changed to the Wigner-Seitz estimate $C_\mathrm{M}=-9/10$ \citep{1961ApJ...134..669S} through a command-line option. Other lattice structures can easily be implemented if needed by providing the suitable value for $C_\mathrm{M}$. 

\subsection{Accuracy and robustness}

For illustration, the outer crust composition of a non-accreting magnetar with a magnetic field $B_* = 100$ was computed. For comparison with the previous studies of \citet{2020PhRvC.101c2801C} and \citet{2020PhRvC.101f5802C}, atomic masses were taken from the experimental data of the 2016 Atomic Mass Evaluation (AME2016; \citealt{2017ChPhC..41c0002H, 2017ChPhC..41c0003W}) and supplemented by the HFB-27 model \citep{2013PhRvC..88f1302G} based on the Hartree–Fock–Bogoliubov method. Experimental measurements of copper isotopes from \citet{2017PhRvL.119s2502W} were also incorporated. The masses $M'(A,Z)$ were then inferred from these atomic masses by subtracting the electron binding energy according to Eq. (A4) of \citet{2003RvMP...75.1021L}, while retaining the rest mass contribution of the electrons. 
The resulting outer crust composition is summarized in Table \ref{tab:strat}. 
It remains close to that obtained for an unmagnetized neutron star in \citet{2020PhRvC.101c2801C}. The only difference in the sequence of equilibrium nuclides is the absence of \element[][66]{Ni}, consistent with Table 1 of \citet{2022Univ....8..328C}, where the composition was determined using the traditional minimization approach.
As shown for strongly magnetized magnetars in \citet{2020PhRvC.101f5802C}, the increase in matter density induced by the magnetic field leads to a more uniform distribution of nuclides. As a result, the baryonic content of the shallow layers becomes closer to that of the deeper layers compared to an unmagnetized neutron-star crust. The contrast nevertheless remains more pronounced here than in the strongly magnetized case.
The main difference between the three compositions is the number of Landau-Rabi levels populated by electrons, as $\nu_\mathrm{max} \rightarrow \infty$ for all layers in the ordinary neutron star and $\nu_\mathrm{max} = 0$ for the strongly magnetized magnetar. 

To assess the accuracy of the approximate equilibrium conditions, Eqs. (\ref{eq:layer_transition}) and (\ref{eq:layer_transition_y1y2}), the outer crust composition of the same magnetar was computed by solving the exact conditions \eqref{eq:gibbs_cond} and \eqref{eq:pressure_cond}. The relative deviations between the exact crustal properties and those computed with the approximate equations are listed in Table \ref{tab:errors}. Focusing on the layers with $\nu_\mathrm{max}>0$, the deviations for the transition pressure and nucleon number densities are typically on the order of $10^{-3}\%$, or less. They reach $\sim 0.2\%$ at most. The deviations for the parameter $x_\mathrm{e}$ do not exceed $\sim 0.06\%$, as do those for the relative depths. Some of the relative abundance deviations are higher, reaching $1.5\%$. This is expected as they are determined from pressure differences. Finally, the very low deviations for the baryon chemical potential, below $4 \times 10^{-4}\%$, ensure that the sequence of equilibrium nuclides is correctly reproduced. 

The robustness of the program was assessed by systematically comparing the resulting sequences of equilibrium nuclides with those obtained by solving the exact equilibrium conditions, using 17 nuclear mass tables to complement the 2020 Atomic Mass Evaluation \citep{2021ChPhC..45c0002H, 2021ChPhC..45c0003W}: 
HFB-24 \citep{2013PhRvC..88b4308G}, 
HFB-27 \citep{2013PhRvC..88f1302G}, 
HFB-31 \citep{2016PhRvC..93c4337G}, 
BSkG4 \citep{2025EPJA...61...35G}, 
BML \citep{2022PhRvC.106b1303N}, 
FRDM \citep{2016ADNDT.109....1M}, 
DRHBc, with both even-even \citep{2022ADNDT.14401488Z} and even-Z \citep{2024ADNDT.15801661G} nuclei,  
DD-ME2, DD-ME$\delta$, DD-PC1, NL3* \citep{2014PhRvC..89e4320A, 2015PhRvC..91a4324A, 2016PhRvC..93e4310A}, 
SkP, SkM*, SLy4, SVmin, and UNEDF1 \citep{2012Natur.486..509E}. 
For each mass table, calculations were performed for all integer values of $B_*$ between 1 and 1300. Above this threshold, the entire crust lies in the strongly quantizing regime \citep[see][]{2020PhRvC.101f5802C}.

\begin{table}[!b]
\caption{Differences between the approximate and exact sequences of equilibrium nuclides.} 
\label{tab:differences}   
\renewcommand{\arraystretch}{1.25}
\centering                          
\begin{tabular}{>{\centering\arraybackslash}p{0.24\columnwidth} >{\centering\arraybackslash}p{0.31\columnwidth} >{\centering\arraybackslash}p{0.31\columnwidth}}  
\hline\hline 
Mass table & Nuclide & Magnetic field ($B_*$) \\
\hline
\rule[0em]{0pt}{1.1em}HFB-27 & \element[][80]{Ni} ($+$) & $[317,320]$ \\ 
BSkG4 & \element[][126]{Ru} ($-$) & $[648,649]$ \\
SLy4 & \element[][124]{Mo} ($-$) & $[808,809]$ \\
SkM* & \element[][120]{Sr} ($+$) & 1015 \\
DRHBc & \element[][134]{Se} ($+$) & 335 \\
      & \element[][140]{Sr} ($-$) & 673 \\
\hline
\end{tabular}
\tablefoot{
The listed nuclides are either present ($+$) or absent ($-$) from the approximate sequence of equilibrium nuclides relative to the exact sequence for the indicated magnetic field strengths.
}
\end{table}

For 11 out of the 17 mass tables considered, the approximate and exact calculations yield identical sequences of equilibrium nuclides. For the remaining tables, only minor differences are found. These fall into two categories: (i) a nuclide appears in the approximate sequence over a limited range of magnetic field strengths while being absent from the exact sequence, and (ii) a nuclide is missing from the approximate sequence where it is present in the exact one. The cases identified for fields $B_*\in [1,1300]$ are summarized in Table \ref{tab:differences}. Overall, these discrepancies between approximate and exact calculations remain very limited. They typically occur over very narrow intervals of magnetic-field strengths and involve at most two nuclides. Out of the 22100 compositions computed, only 11 sequences of equilibrium nuclei exhibit such differences, the remaining being exactly reproduced.

To evaluate the validity of the Gibbs free energy expansion to first order in $\alpha$, we tracked the relative density discontinuity $|\delta n_\mathrm{e}|/n_\mathrm{e} \equiv |n_\mathrm{e,2} - n_\mathrm{e,1}|/n_\mathrm{e,1}$ for all candidate transitions. For each of the 17 mass tables and 1300 magnetic field values considered, fewer than $0.2\%$ of all computed transitions exhibit $|\delta n_\mathrm{e}|/n_\mathrm{e} \geq 0.1$. The first-order expansion in $\alpha$ therefore remains well justified in the vast majority of cases.

Finally, we analyzed the timing performance of our program in comparison with the traditional minimization method, using the exact expressions (\ref{eq:electron_density}) and (\ref{eq:electron_pressure}) to compute the electron properties. Using an Intel Core i7-8086K processor, the calculation of the outer crust composition with HFB-27 and $B_*=100$ takes about 0.20~s with our iterative approach. Considering a pressure grid of about 20000 points, going from $P=10^{-12}~\mathrm{MeV~fm}^{-3}$ to $P=P_\mathrm{drip}$ with a pressure step of $\delta P = 10^{-3}P$ (to ensure errors of the same order as those for the iterative approach in most layers), the traditional method, not including the abundance and relative depth calculations, takes 591.46~s, i.e., $3\times10^3$ times longer. 

\begin{figure}[!t]
    \includegraphics[width=\hsize-0.5cm]{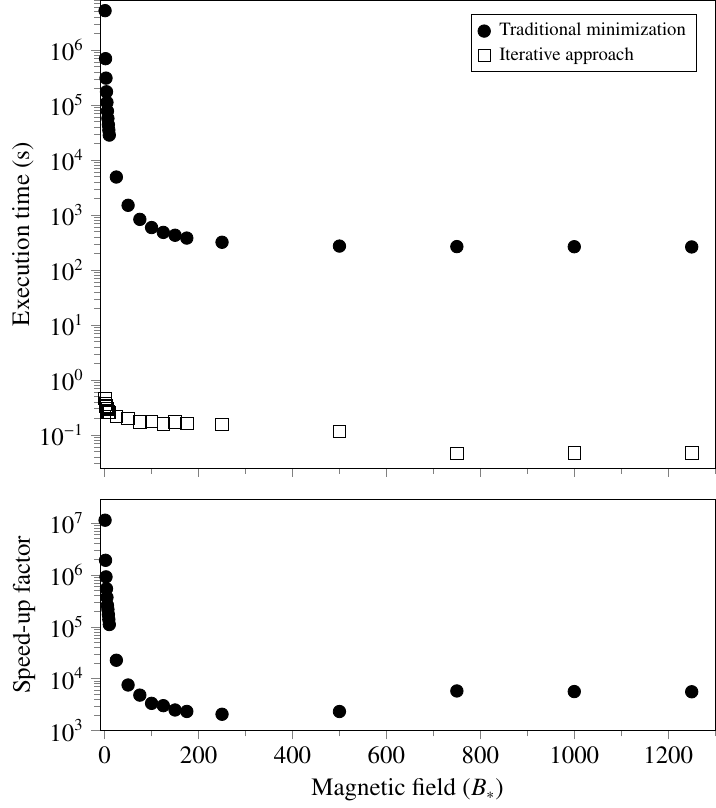}
    \caption{Execution time (top) and corresponding speed-up factor (bottom) as a function of the magnetic-field strength. Each point corresponds to an independent outer-crust calculation. The speed-up factor is defined as the ratio of the execution time of the traditional minimization to that of the iterative approach.}
     \label{fig:timing}
\end{figure}

As shown in the top panel of Fig.~\ref{fig:timing}, the computation time strongly depends on the magnetic field. It is highest for very small values of $B_*$, notably for $B_*\lesssim15$, where hundreds to over a thousand Landau-Rabi levels are populated in the deepest layers of the outer crust. The typical surface magnetic fields observed in magnetars correspond to $B_* \approx 10-100$, where our program is particularly efficient. As shown in the bottom panel of Fig.~\ref{fig:timing}, the speed-up factor between the traditional method and the iterative approach reaches $10^5$ for $B_*=10$.

This gain in performance originates from both the iterative scheme and the use of the expansions (\ref{eq:electron_dens_w}) and (\ref{eq:pressure_w}) for the electron properties. Nevertheless, the repeated evaluation of Hurwitz zeta functions still introduces a non-negligible computational overhead. As a result, our program is slightly slower in regimes where transitions with $\nu_\mathrm{max} \geq 2$ occur. This explains the sharp decrease in execution time observed in Fig.~\ref{fig:timing} between $B_*=500$, where such transitions still occur, and $B_*=750$, where all transitions satisfy $\nu_\mathrm{max}\leq 1$. Over the full range $B_*=1-1300$, our program nonetheless remains $2\times10^3$ to $10^{7}$ times faster than the traditional approach. Still using the HFB-27 mass table, compositions for every integer magnetic field value between $B_*=1$ and $1300$ were computed in only 129.19~s (2.15 minutes). For comparison, this computing time is estimated to more than 80 days with the traditional method.
This corresponds to a total speed-up factor of $5 \times 10^4$. This computational gain is on the same order as that obtained with the previous programs \citep{2020PhRvC.101c2801C, 2020PhRvC.101f5802C} in the limits of unmagnetized and strongly magnetized crusts for the same pressure step and range. Similar performance was obtained for all mass tables tested, with total runtimes ranging from one to three minutes depending on the number of nuclides included in the model.

Our program is therefore particularly well-suited for systematic calculations of the outer crust composition of neutron stars across a wide range of magnetic-field strengths and nuclear mass models. Large-scale statistical studies, such as those performed in \citet{2020PhRvC.101c5804P} and \citet{universe7050131} to assess the impact of nuclear mass uncertainties on the outer-crust composition of unmagnetized neutron stars, require several tens of thousands of nuclear mass tables. In the case of magnetars, the additional sampling over magnetic-field strengths renders traditional approaches computationally prohibitive.

\section{Conclusions}

The iterative method previously developed to determine the structure of the outer crust of unmagnetized \citep{2020PhRvC.101c2801C} and strongly magnetized neutron stars \citep{2020PhRvC.101f5802C} has been extended to arbitrary magnetic-field strengths in the program \texttt{magcrust}. The generalized scheme consistently accounts for Landau-Rabi quantization of electron motion across the full range of magnetic-field strengths relevant for magnetars. Transitions between adjacent crustal layers are computed using approximate equilibrium conditions, with analytic solutions in the limits of unmagnetized matter and strongly quantizing magnetic fields. In the weakly quantizing regime, solutions are 
evaluated numerically using approximate expansions for electron properties.
The method is robust across a wide range of nuclear mass models and magnetic-field strengths. Relative deviations with exact crustal properties do not exceed a few percent, and discrepancies in the sequence of equilibrium nuclides were found in only $0.05\%$ of all computed compositions. 
The generalized scheme, furthermore, remains significantly less computationally demanding than the traditional minimization procedure, with speed-up factors of $10^3-10^7$ for magnetic-field strengths between $B_*=1$ and $1300$. 

Overall, this new program provides an efficient and reliable tool for modeling the composition and structure of neutron-star outer crusts over the full range of magnetic-field strengths relevant for astrophysical applications. As such, it opens the door for sensitivity analyses requiring calculations over a very large number of nuclear mass tables, such as those performed for unmagnetized neutron-star crusts. The present implementation is restricted to zero temperature. An extension to include finite-temperature effects for determining the equilibrium composition of hot young magnetars and the crystallization of their crust is left for future work.

\section*{Data availability}
The developed computer program and full results are freely available in \citet{servais_2026_20085135}, at \url{https://doi.org/10.5281/zenodo.20085135}.

\begin{acknowledgements}
      NC thanks Zhivko Stoyanov and Konstantin Shegunov for discussions during the early stage of this work. 
      JS is a FRIA grantee of the Fonds de la Recherche Scientifique - FNRS (Communauté française de Belgique).
\end{acknowledgements}

\raggedbottom
\bibliographystyle{aa} 
\bibliography{main.bib} 

\end{document}